\documentclass[%
 reprint,
 amsmath,amssymb,
 aps
]{revtex4-2}
\usepackage{amsmath}
\usepackage{amsfonts}
\usepackage{amssymb}
\usepackage{subfigure}
\usepackage{graphicx}
\usepackage{dcolumn}
\usepackage{bm}
\usepackage{acronym}
\newacro{FM}{ferromagnet}
\newacro{HM}{heavy-metal}
\newacro{DMI}{Dzyaloshinskii-Moriya interaction}
\newacro{CIP}{current-in-plane}
\newacro{CPP}{current-perpendicular-to-plane}
\newacro{PMA}{perpendicular magnetic anisotropy}
\newacro{AFM}{anti-ferromagnetic}
\newacro{LIF}{leaky-integrate-and-fire}
\newacro{LTD}{long-term depression}
\newacro{LTP}{long-term potentiation}
\newacro{SNN}{spiking neural network}
\newacro{TL}{top layer}
\newacro{BL}{bottom layer}
\newacro{OOMMF}{Object-Oriented MicroMagnetic Framework}
\newacro{MTJ}{magnetic tunnel junction}
\newacro{CIM}{compute-in-memory}

\usepackage[margin=0.54in]{geometry}

\usepackage{hyperref}

\begin{document}


\title{Bilayer-skyrmion-based design of neuron and synapse for spiking neural network}

\author{Debasis Das}
\altaffiliation{eledd@nus.edu.sg}
\author{Yunuo Cen}
\author{Jianze Wang}
\author{Xuanyao Fong}
\altaffiliation{kelvin.xy.fong@nus.edu.sg}
\affiliation{Department of Electrical and Computer Engineering, National University of Singapore, Singapore 117583}%

\date{\today}

\begin{abstract}
Magnetic skyrmion technology is promising for the next-generation spintronics-based memory and neuromorphic computing due to their small size, non-volatility and low depinning current density.
However, the Magnus force originating from the skyrmion Hall effect causes the skyrmion to move along a curved trajectory, which may lead to the annihilation of the skyrmion in a nanotrack during current-induced skyrmion motion.
Consequently, circuits utilizing skyrmionic motion need to be designed to limit the impact of the skyrmion Hall effect.
In this work, we propose a design of an artificial neuron, and a synapse using the bilayer device consisting of two antiferromagnetically exchange coupled ferromagnetic layers, which achieves robustness against the skyrmion Hall effect by nullifying the Magnus force. Using micromagnetic simulations, we show that the bilayer device can work as an artificial neuron and also as a synapse by modifying its uniaxial anisotropy.
We also demonstrate that our proposed skyrmionic synapse has an intrinsic property of perfectly linear and symmetric weight update, which is highly desirable for the synapse operation.
A spiking neural network implemented using our proposed synapse and neuron was simulated and showed to achieve 96.23\% accuracy in performing classification on the MNIST handwritten digit dataset.
\end{abstract}

\maketitle

\section{Introduction}\label{Intro}
The increasing demands for data-intensive computing requires a paradigm shift in computing approaches to push the envelope of energy efficiency for next-generation of intelligent electronic systems.
Neuromorphic computing~\cite{mead1990neuromorphic,serrano2013stdp} is one promising approach that takes inspiration from the biological brain to design intelligent electronic systems that are energy efficient while being able to learn and to perform cognitive tasks with human-like accuracies.
Many current works \cite{chen2015efficient,jung2022crossbar} focus on leveraging the \ac{CIM} architecture to accelerate the computation of tensor products by overcoming the von~Neumann bottleneck.
However, even lower energy consumption can be achieved by using an event-based computation model such as the \ac{SNN}, which avoids redundant computations (\emph{i.e,} multiplications and additions where at least one operand is zero).
In this work, we demonstrate the design of a skymionic hardware accelerator for the \ac{SNN}, which is different but inspired by the \ac{CIM} architecture.

\begin{figure}[t!]
	\includegraphics[scale=0.3]{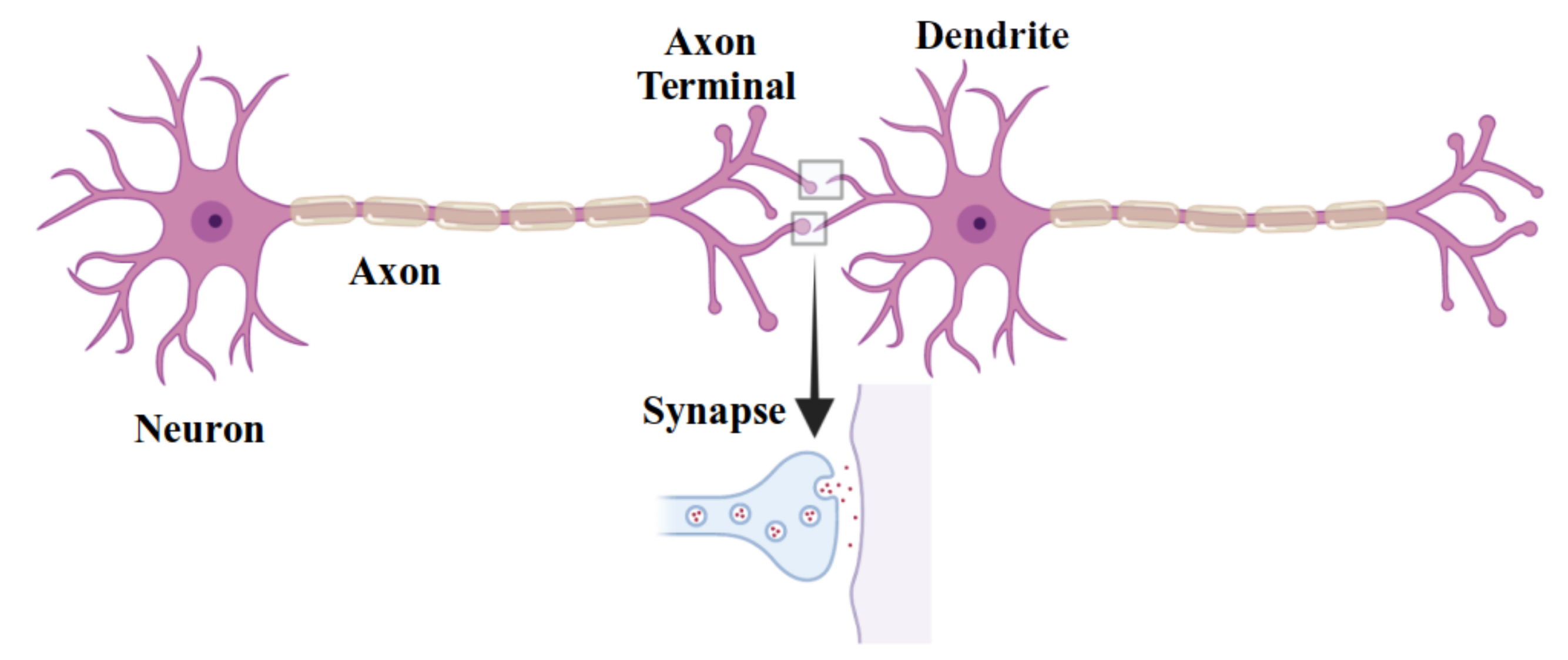}
	\caption{Schematic of biological neurons connected by two synapses.}
	\label{Fig:Bio_neuron_synapse}
\end{figure}

Other than mimicking the signaling mechanisms in the biological brain, research into neuromorphic circuits also attempts to mimic the behavior of biological neural networks \cite{kumar2020third,lotter2020neural}.
Two of the fundamental components of the neural network in the brain are the neuron and synapse, where billions of neurons are connected in a massively parallel fashion via trillions of synapses.
The diagram in Fig.~\ref{Fig:Bio_neuron_synapse} shows two neurons connected by synapses through which neural signals are transmitted from one neuron to another.
Moreover, these signals are modified by the synaptic weights during the transmission \cite{abbott2004synaptic,liu2022analog}.
Proper device design for the artificial neuron and synapse is required to achieve low power consumption in the neuromorphic circuits.

Various emerging devices such as phase-change memory~\cite{tuma2016stochastic,kim2015nvm}, resistive memory~\cite{yu2011electronic,park2012rram,park2013nanoscale}, and spintronics~\cite{sengupta2016proposal,brigner2019shape,zhang2016all,chen2018magnetic} have been explored for implementing the artificial neuron and synapse.
Among these devices, spintronics is arguably the most promising technology due to its non-volatile nature, ultrafast dynamics~\cite{seifert2016efficient}, higher endurance~\cite{finocchio2021promise}, low power~\cite{zhang2014spintronics} and stochastic nature~\cite{shim2017stochastic,das2021fokker}.
Among various spintronic devices, the magnetic skyrmion has gained a lot of attention in recent years due to its small size, ultralow depinning current density and topological stability~\cite{muhlbauer2009skyrmion,sampaio2013nucleation,jiang2015blowing}.

The magnetic skyrmion is a vortex-like spin texture observed in bulk~\cite{jonietz2010spin} as well as in a thin \ac{FM}~\cite{woo2016observation,yu2016room} grown on a \ac{HM}, and stabilized by \ac{DMI}~\cite{moriya1960anisotropic,rohart2013skyrmion} arising from the broken inversion symmetry.
However, the skyrmion Hall effect~\cite{chen2017skyrmion} poses a challenging problem for designing compact high-speed skyrmionic nanotrack devices.
When a spin-polarized current is applied to move the skyrmion in an FM layer, it follows a curved path governed by the skyrmion Hall angle, which is determined by the combined effect of force due to the spin-torque and Magnus force that act parallel and perpendicular to the current direction, respectively~\cite{nagaosa2013topological, sampaio2013nucleation}.
The Magnus force is directly proportional to the skyrmion velocity, which is proportional to the applied current density~\cite{kang2016skyrmion}.
Consequently, a large Magnus force results in a transverse velocity for the skyrmion~\cite{zhang2016magnetic} and may cause it to be displaced beyond the edge of the nanotrack and be accidentally annihilated. Hence, there is an upper limit on the current density that can be applied to the device, which limits its operation speed.

Note also that there are two methods to inject the spin current into the device to move the skyrmion along the nanotrack: (i) the \ac{CIP} scheme and (ii) the \ac{CPP} scheme.
It was shown that for the same current density, the magnetic skyrmion can achieve much higher velocity in the \ac{CPP} scheme compared to the \ac{CIP} scheme~\cite{kang2016skyrmion}.
Thus, to obtain a high-speed operation of skyrmion-based devices, the \ac{CPP} scheme is preferred and is also utilized in our proposed devices.

There have been few reports which emulate neuron and synapse functionality using skyrmions in monolayer devices, which consist of a thin FM layer with \ac{PMA} grown on an \ac{HM}.
Micromagnetic simulations in Ref.~\cite{li2017magnetic}, demonstrate a skyrmion-based artificial neuron by injecting spin current via the \ac{CIP} scheme, where the Magnus force is forcefully nullified when the non-adiabatic spin-transfer-torque factor, $\beta$, is exactly equal to the Gilbert damping constant, $\alpha$.
This method is sensitive to the material parameters which are difficult to control precisely.
Furthermore, the artificial synapse device needs to have well-controlled and highly symmetrical synaptic weights for \ac{LTP} and \ac{LTD}~\cite{fu2019mitigating,xia2019memristive}. 
Previous works~\cite{huang2017magnetic,bhattacharya2019low,chen2020nanoscale} propose an anisotropy barrier region in the middle of the nanotrack which cause skyrmions to experience a force when they are in the vicinity of the barrier while crossing from presynapse to postsynapse region, or vice versa.
Due to the presence of the Magnus force in these devices, the motion of the individual skyrmions cannot be precisely controlled.
This results in a non-linear relationship between the synaptic weights and the number of pulses applied to move the skyrmions, which can make the synaptic device non-linear and asymmetric in nature.
There also remains the possibility of skyrmion annihilation due to the Magnus force, which can affect its functionality.

To alleviate the issue of skyrmion annihilation due to the Magnus force, we propose to nullify the Magnus force by utilizing \ac{AFM} exchange coupling in a bilayer system consisting of two \ac{FM} nanotracks with opposite magnetization~\cite{zhang2016magnetic}.
With sufficiently large \ac{AFM} exchange coupling between the \ac{FM} layers, the Magnus force acting on each skyrmion in the corresponding layer is exactly canceled.
Moreover, the skyrmion can move along a perfectly straight line in the direction of the applied current.
The absence of the Magnus force stabilizes the device operation by eliminating accidental skyrmion annihilation and also enhances the upper limit of the current density significantly.
Thus, the skyrmion velocity can be significantly increased to achieve high-speed operation of the skyrmionic devices. 
\begin{figure}[t!]
	\includegraphics[scale=0.27]{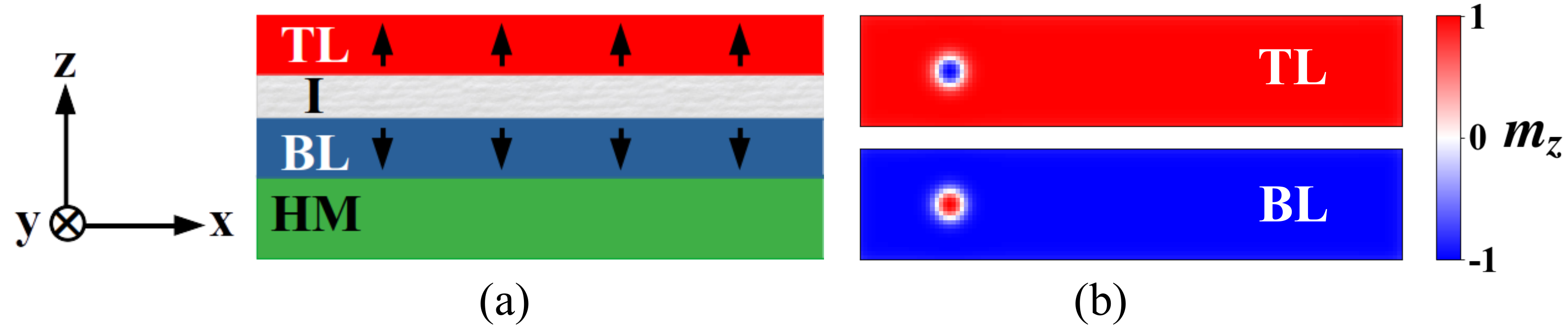}
	\caption{(a) Side-view of the schematic of bilayer device. Arrows in the TL and BL show magnetization direction. (b) Two-dimensional color plot for $m_z$ on the $x\mbox{-}y$ plane for TL and BL.}
	\label{Fig:device}
\end{figure}

In this work, we use micromagnetic simulations to demonstrate the operation of the proposed skyrmion-based bilayer devices as an artificial neuron as well as an artificial synapse. We study the leaky integrate-and-fire (LIF) functionality of the proposed artificial neurons by introducing an anisotropy gradient in the FM layer of the nanotrack. A comparison between the monolayer and bilayer device schemes is also performed. We also show that, by modifying the anisotropy constant profile, a similar device structure with a longer length can be used as a synapse, which indicates the ability to implement a complete neuromorphic system using the proposed skyrmion-based devices. A spiking neural network (SNN) implemented using the proposed device was simulated to demonstrate the functionality. The performance of the SNN on classifying the MNIST handwritten digit dataset using a modified spike-based backpropagation training approach~\cite{lee2020enabling} is evaluated.

\section{Device Concept \& Model}\label{formulation}

In this section, we first describe the device structure and how it is used to implement the functionality of the neuron and synapse in Sec.~\ref{Sec:device}.
Thereafter, in Sec.~\ref{Sec:math_model}, the micromagnetic model of the device is presented.

\subsection{Device structure}\label{Sec:device}

In a bilayer system, two FM layers (namely top layer, TL, and bottom layer, BL) are separated by a thin insulator (I) layer, and the BL layer is in contact with the bottom HM layer (which gives rise to DMI in the device) as shown in Fig.~\ref{Fig:device}(a).
Both the FM layers are assumed to have perpendicular magnetic anisotropy (PMA).
The magnetization for the TL and BL are assumed to be along $+z$ and $-z$ direction, respectively, and are coupled by AFM exchange interaction.

To nucleate a skyrmion in the bilayer system, a 20~ps current pulse of magnitude $5\times{}10^{10}~ \mathrm{A/cm^2}$ that is spin-polarized along the $-z$-direction is injected locally at the desired location of the TL~\cite{zhang2016magnetic} followed by 180~ps of relaxation time.
The skyrmion nucleated in the TL by this current has $-z$-directed magnetization at its center.
Due to AFM coupling, nucleation of the skyrmion in TL also nucleates a skyrmion with $+z$-directed magnetization at its center in BL, just below the skyrmion in TL.
The skyrmions nucleated in TL and BL are shown in Fig.~\ref{Fig:device}(b), where the $z$-component of the normalized magnetization (\emph{i.e.}, $m_{z}$ in the $x\mbox{-}y$ plane) is plotted and the color bar shows its magnitude.
The parameters for the simulation are described Sec.~\ref{Sec:results}.

To move the skyrmion, an electrical charge current is passed through HM in the $x$-direction, which gives rise to a spin-polarized current that flows in the $+z$-direction, having spins that are oriented along the $y$-direction.
In combination with the AFM coupling between BL and TL, this spin-polarized current moves the skyrmions in BL and TL along the nanotrack.
The motion of the skyrmion in the TL for this bilayer device may be used to emulate the functionality of neurons and synapses.

The widely used LIF model is a bio-plausible model for the spiking behavior of a biological neuron. The presence of incoming excitatory stimulus increases the membrane potential, which is known as ``integration''.
In the absence of any stimulus, the membrane potential of the neuron gradually returns to a rest potential, which is called as ``leak'' because it is akin to leakage current gradually discharging a capacitor to a steady-state voltage.
If the membrane potential exceeds a threshold during the integration, the neuron ``fires'' and produces an output spike.
Thereafter, it enters a refractory period where its membrane potential is reset to the rest potential and the neuron is unresponsive to input spikes until the refractory period has elapsed.
In the following, we discuss how the device in Fig.~\ref{Fig:device} can be designed to emulate the behaviour of a LIF neuron.

\begin{figure}[t]
	\includegraphics[scale=0.18]{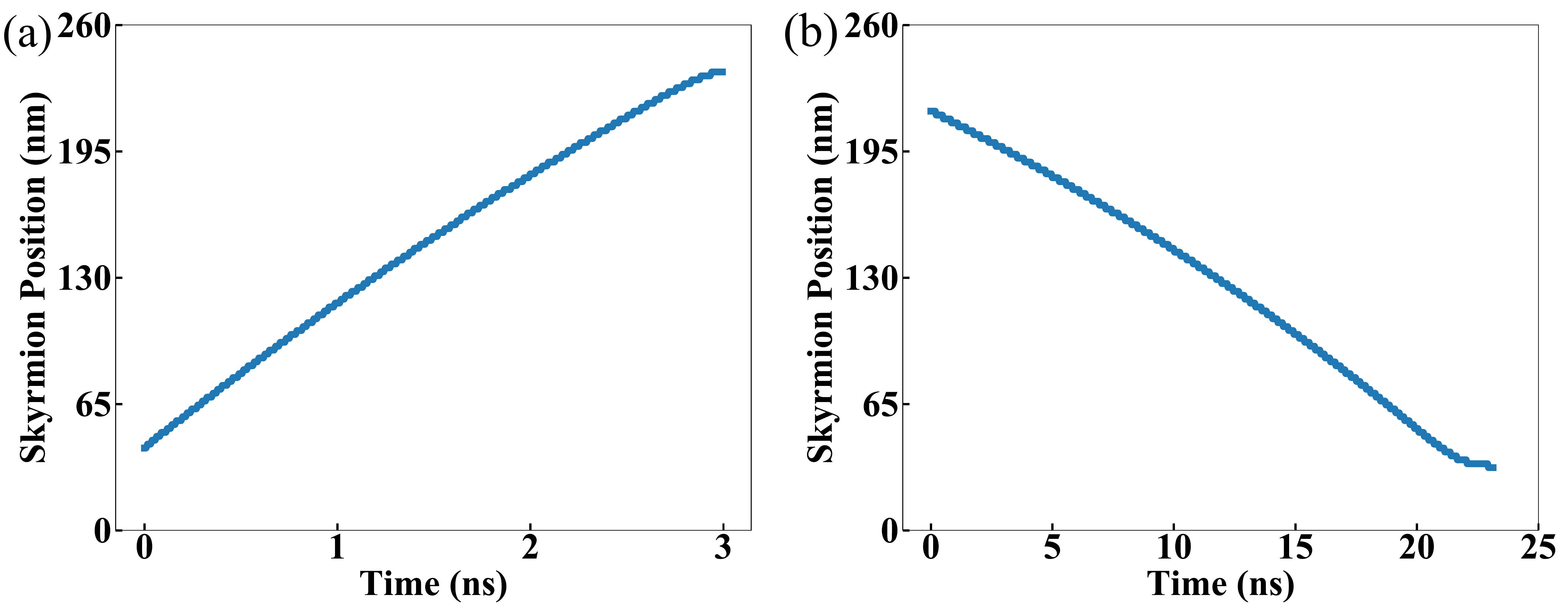}
	\caption{Motion of the skyrmion under the influence of anisotropy gradient in BL. The plot of the TL skyrmion center (along the length of the nanotrack) with time for (a) constant current density of 30 $\mathrm{MA/cm^2}$, demonstrates the integration process, (b) anisotropy gradient only (with no external current), demonstrating the leaky functionality.}
	\label{Fig:Leak_Integration}
\end{figure}

First, consider if the BL in the structure in Fig.~\ref{Fig:device} has a constant uniaxial anisotropy gradient that increases toward the $+x$-direction whereas the uniaxial anisotropy in the TL is constant everywhere.
Assume that the device is 260~nm long and the left edge of the device is at $x=0~\text{nm}$.
If a skyrmion-pair is nucleated in the TL and BL (Fig. \ref{Fig:device}(b)) at $x=50~\text{nm}$ and a constant 30~MA/cm$^{2}$ charge current density is passed through the HM along the $x$-direction, the skyrmions in BL and TL will move in unison along the nanotrack in the $+x$-direction.
The position of the skyrmion center in the TL versus time is shown in Fig.~\ref{Fig:Leak_Integration}(a).
Due to the coupled motion, the skyrmion in the BL shows the exact same behaviour (figure not shown).
If the position of the skyrmion is considered as the membrane potential of the neuron, the behaviour exhibited in Fig.~\ref{Fig:Leak_Integration}(a) is analogous to the time-integration of the input charge current.
In this example, the large charge current density is able to move the skyrmion across the length of the nanotrack in about 3~ns.

Now consider if the skyrmion is at $x=225~\text{nm}$ and no charge current flows through the device.
It has been reported that if a skyrmion is placed in an anisotropy gradient, it can move due to this anisotropy gradient even in the absence of any applied current~\cite{das2019skyrmion}.
Therefore, the skyrmion in the BL will gradually move in the $-x$-direction.
Due to the AFM exchange coupling, the skyrmion in TL will move in unison with that in BL.
As shown in Fig.~\ref{Fig:Leak_Integration}(b), skyrmions will move until $x=25~\text{nm}$ and stop when the effect of the anisotropy gradient is exactly cancelled by the demagnetizing field due to the edge of the device.
This behaviour is analogous to the leak of the LIF neuron. Hence, in our proposed device, the position of the skyrmion in TL mimics the membrane potential of a neuron.
In this example, the leak is a slow process and it takes about 23~ns for the skyrmion to traverse the length of the nanotrack due only to the anisotropy gradient.

\begin{figure}[t]
	\includegraphics[scale=0.2]{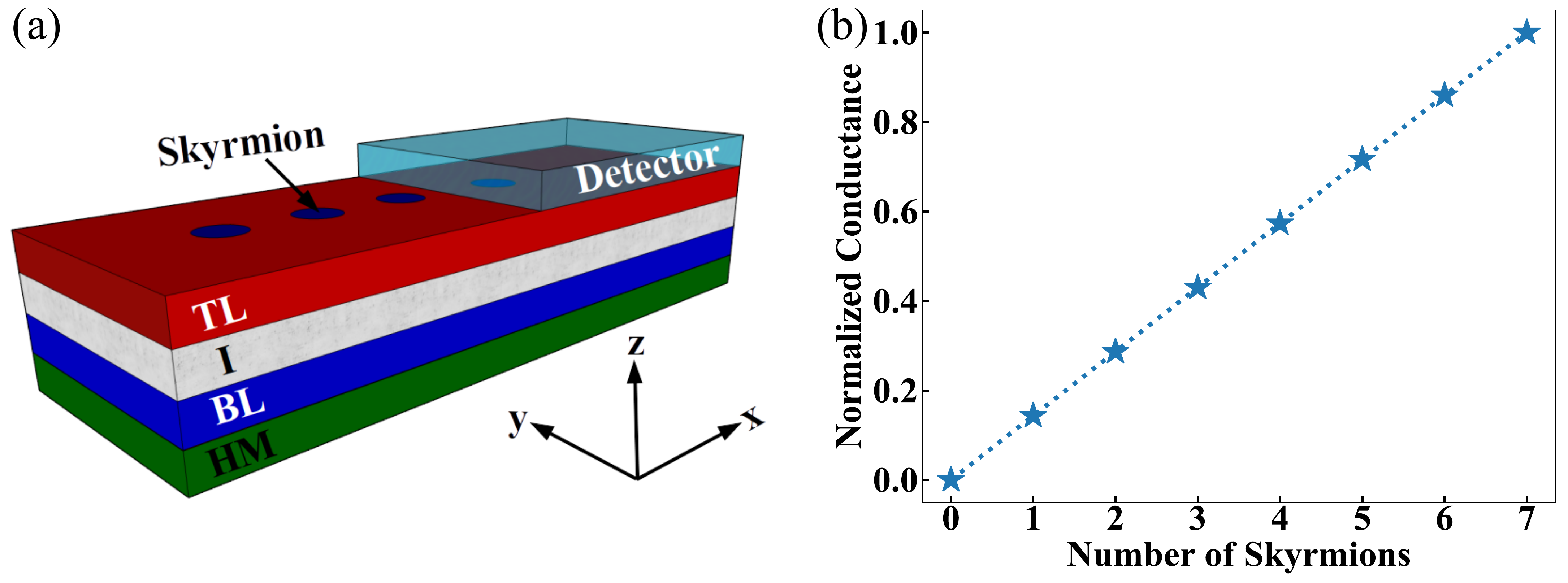}
	\caption{(a) Schematic of a 3D view of the synaptic device, with the skyrmions(blue circle) shown at TL and the detector at the right side of the TL nanotrack. (b) Plot for normalized conductance vs. the number of skyrmions in the detector.}
	\label{Fig:Synapse_Demo}
\end{figure}

Unlike neurons, synapses have neither ``leak'', ``integrate'' nor ``fire'' behaviour.
Instead, they need to emulate different weight values, which modulates the neuronal signal passing from the pre-synaptic neuron (or \emph{pre-neuron}) to post-synaptic neuron (or \emph{post-neuron}) through it.
Thus, to implement the synapse functionality in the device in Fig.~\ref{Fig:device}, the anisotropy is uniform in both TL and BL (\emph{i.e.}, no anisotropy gradient in BL).
Next, the TL nanotrack is divided into two equal portions along its length.
A detector unit is placed at top on the right side of the nanotrack as shown in Fig.~\ref{Fig:Synapse_Demo}(a).
Instead of a single skyrmion in the nanotrack, multiple skyrmions are nucleated at the left side of the nanotrack.
If the device is driven by a current along the $+x$-direction, the skyrmions will move into the right half of the nanotrack one by one.
The presence of a skyrmion in the detector region changes the conductance of this region, which can be measured electrically through the change in the magneto-resistance (MR) \cite{hanneken2015electrical, crum2015perpendicular, huang2017magnetic}. 
The conductance measured by the detector on the right-side changes linearly with the number of skyrmions in the TL underneath it.
Hence, as the skyrmions move into that region one by one, the conductance through the detector increases as illustrated in Fig.~\ref{Fig:Synapse_Demo}(b).
Note that the conductance plotted in Fig.~\ref{Fig:Synapse_Demo}(b) is normalized to the total number of skyrmions.
Hence, the different conductance values may be used to represent the synaptic weight values that the synapse needs.
Thus, by controlling the number of skyrmions present under the detector unit, the weight values of the synaptic device can be manipulated.

Before discussing the detailed operation of our proposed device as neuron and synapse, we first discuss the model used in our micromagnetic simulations of the devices.

\subsection{Micromagnetic model}\label{Sec:math_model}

The Hamiltonian for each layer $l$ (=TL, BL) is given by\begin{equation}
	\begin{split}
		H^l=&K_u^l\sum_i\left(1-\left(\mathbf{m}_{i}^{l}\cdot\mathbf{z}\right)^2\right)+\sum_{\langle{}i,j\rangle}\mathbf{D}\cdot\left(\mathbf{m}_i^l \times \mathbf{m}_j^l\right) \\
		&+ H_\text{dipole}^{l} + A_\text{intra}\sum_{\langle{}i,j\rangle}\mathbf{m}_i^l \cdot \frac{\mathbf{m}_i^l - \mathbf{m}_j^l}{\Delta_{ij}} .
		\label{Eq:Hamiltonian}
	\end{split}
\end{equation}$\mathbf{m}_i^l$ represents the local magnetization at site $i$ for layer $l$, and $\langle{}i,j\rangle$ corresponding to the nearest neighbor interaction in each layer $l$.
The first term in Eq.~(\ref{Eq:Hamiltonian}) represents the perpendicular magnetic anisotropy (PMA) energy with anisotropy constant $K_{u}^{l}$ for the layer $l$.
The second term represents the energy for the DMI with DM vector $\mathbf{D}$, with magnitude given by the DMI constant, $D$.
The third term represents the energy due to dipole-dipole interaction, and the fourth term represents intra-layer exchange interaction with exchange stiffness constant, $A_\text{intra}$. $\Delta_{ij}$ is the discretization size between cell~$i$ and cell~$j$.
Apart from $H^{TL}$ and $H^{BL}$, there is also an energy term that represents the interlayer AFM exchange interaction between the TL and the BL, which is given by \begin{equation}
	H_\text{inter}=\sigma \sum_k \frac{1-\mathbf{m}_k^{TL} \cdot \mathbf{m}_k^{BL}}{\Delta_k}
	\label{Eq:Exchange_energy}
\end{equation}$\sigma$ is the bilinear surface exchange coefficients between the two layers.
$\Delta_{k}$ is the discretization size in the direction from cell~$k$ of TL to cell~$k$ of BL.
Thus, the net Hamiltonian for the bilayer system is $H_\text{net}=H^{TL}+H^{BL}+H_\text{inter}$, which provides the effective magnetic field, $\mathbf{H}_\text{eff}^{l}=-\partial{H_\text{net}}/\partial \mathbf{m}^{l}$, acting on layer $l$ in the bilayer system.
Dynamics of skyrmions in layer $l$ is simulated by solving the Landau-Lifshitz-Gilbert-Slonczewski (LLGS) equation under the influence of $\mathbf{H}_\text{eff}^{l}$, which is given by~\cite{sampaio2013nucleation,OOMMF}\begin{equation}
	\begin{split}
		\frac{d\mathbf{m}^l}{dt}=&-\gamma \left(\mathbf{m}^l\times \mathbf{H}_\text{eff}^l\right)+\alpha \left(\mathbf{m}^l\times\frac{d\mathbf{m}^l}{dt}\right) \\
		&+\gamma \beta \left(\mathbf{m}^l\times \mathbf{m}_\text{p} \times \mathbf{m}^l\right) \label{LLGS_eq}
	\end{split}
\end{equation}$\gamma=2.211\times{}10^{5}~\mathrm{m/(A\cdot s)}$ is the gyromagnetic ratio, and $\alpha$ is the Gilbert damping factor.
The first and the second terms in Eq. (\ref{LLGS_eq}) are the precession and damping terms, respectively.
The third term in Eq.~(\ref{LLGS_eq}), represents the current-induced spin-torque where $\mathbf{m}_\text{p}$ denotes the spin polarization direction, and $\beta=\hbar{}PJ/(2\mu_{0}e t_\text{FM}M_\text{s})$.
$\hbar$, $P$, $J$, $\mu_{0}$, $e$, $t_\text{FM}$ and $M_\text{s}$ are the reduced Planck's constant, the spin polarization factor, the charge current density, the permeability of free space, the electronic charge, the thickness of the FM layer, and the saturation magnetization of the FM layer, respectively.

\section{Results \& Discussion}\label{Sec:results}

In the following sections, we discuss the emulation of neuron and synapse functionality using a skyrmion-based bilayer system, and evaluate a circuit based on these devices when it emulates an SNN.
The dynamics of skyrmions in TL and BL is simulated with the aid of the DMI extension module~\cite{DMI_extension} in the Object Oriented MicroMagnetic Framework (OOMMF)~\cite{OOMMF}.
The thickness of both layers are assumed to be 1~nm, and the lateral dimensions of these layers in the neuron and synapse devices are 260~nm$\times$50~nm and 1000~nm$\times$50~nm, respectively.
In all our simulations, the simulation volume is discretized into 2~nm$\times$2~nm$\times$1~nm cells.
The simulation parameters are given in Table \ref{tab:table1} and are used for the rest of this work, unless otherwise stated.
The material parameters assumed are those for $\mathrm{Co\vert Pt}$ system~\cite{sampaio2013nucleation, zhang2016magnetic}.

\begin{table}[t]
	\caption{\label{tab:table1}%
		Device simulation parametrs
	}
	\begin{ruledtabular}		
		\begin{tabular}{cc}			
			\textrm{Parameters} & \textrm{Value}\\
			\colrule
			Saturation magnetization, $M_s$ & 580 kA/m  \\ 
			Gilbert damping factor, $\alpha$ & 0.3 \\
			Spin polarization factor, $P$ & 0.4 \\
			PMA constant, $K_u$ & 0.8 $\mathrm{MJ/m^3}$ \\
			Anisotropy gradient, $\Delta K_u$ & 1.538 $\mathrm{GJ/m^4}$\\
			DMI constant, $D$ & 3 $\mathrm{mJ/m^2}$ \\
			Exchange stiffness constant, $A_{intra}$ & 15 pJ/m \\
			Bilinear surface exchange coefficient, $\sigma$ & -6 $\mathrm{mJ/m^2}$ \\
		\end{tabular}
	\end{ruledtabular}
\end{table}

\subsection{The structure as an artificial neuron}\label{Sec:Neuron}

\begin{figure}[t]
	\includegraphics[scale=0.38]{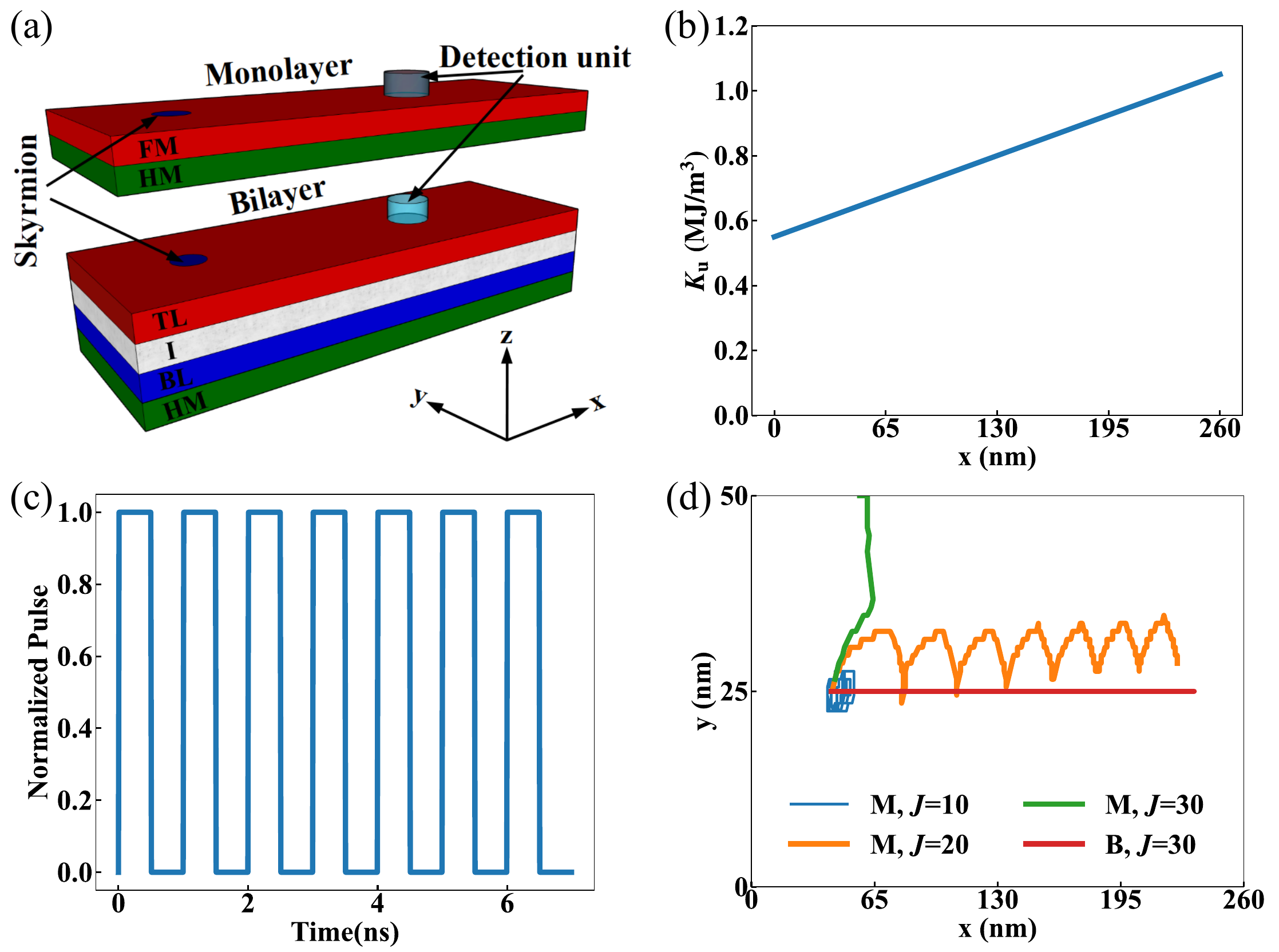}
	\caption{(a) Schematic of the 3D view of the monolayer and bilayer device. (b) Variation of PMA constant $K_\text{u}$ with position along the length of the FM nanotrack for monolayer device. (c) Plot for normalized square pulse train with time, which is given as an input to monolayer as well as bilayer neuron. (d) Locus of the skyrmion center in the nanotrack for monolayer(M) and bilayer(B) devices. $J$ is the current density in $\mathrm{MA/cm^2}$.}
	\label{Fig:monolayer-bilayer_comparison}
\end{figure}

To investigate the effect of the Magnus force on the skyrmion motion in both monolayer and bilayer devices, we simulated both devices.
The monolayer device consists of an FM layer with PMA in contact with an HM layer as shown in Fig.~\ref{Fig:monolayer-bilayer_comparison}(a).
The magnetization of the FM layer is assumed to be along the $+z$-direction. To imitate the leaky functionality of neurons, a constant uniaxial anisotropy gradient in the FM layer of the monolayer device where $K_\text{u}$ increases towards $+x$-direction is assumed.
For the bilayer device, this gradient is incorporated in the BL only, whereas the TL has a constant $K_\text{u}$.

\begin{figure}[t]
	\includegraphics[scale=0.36]{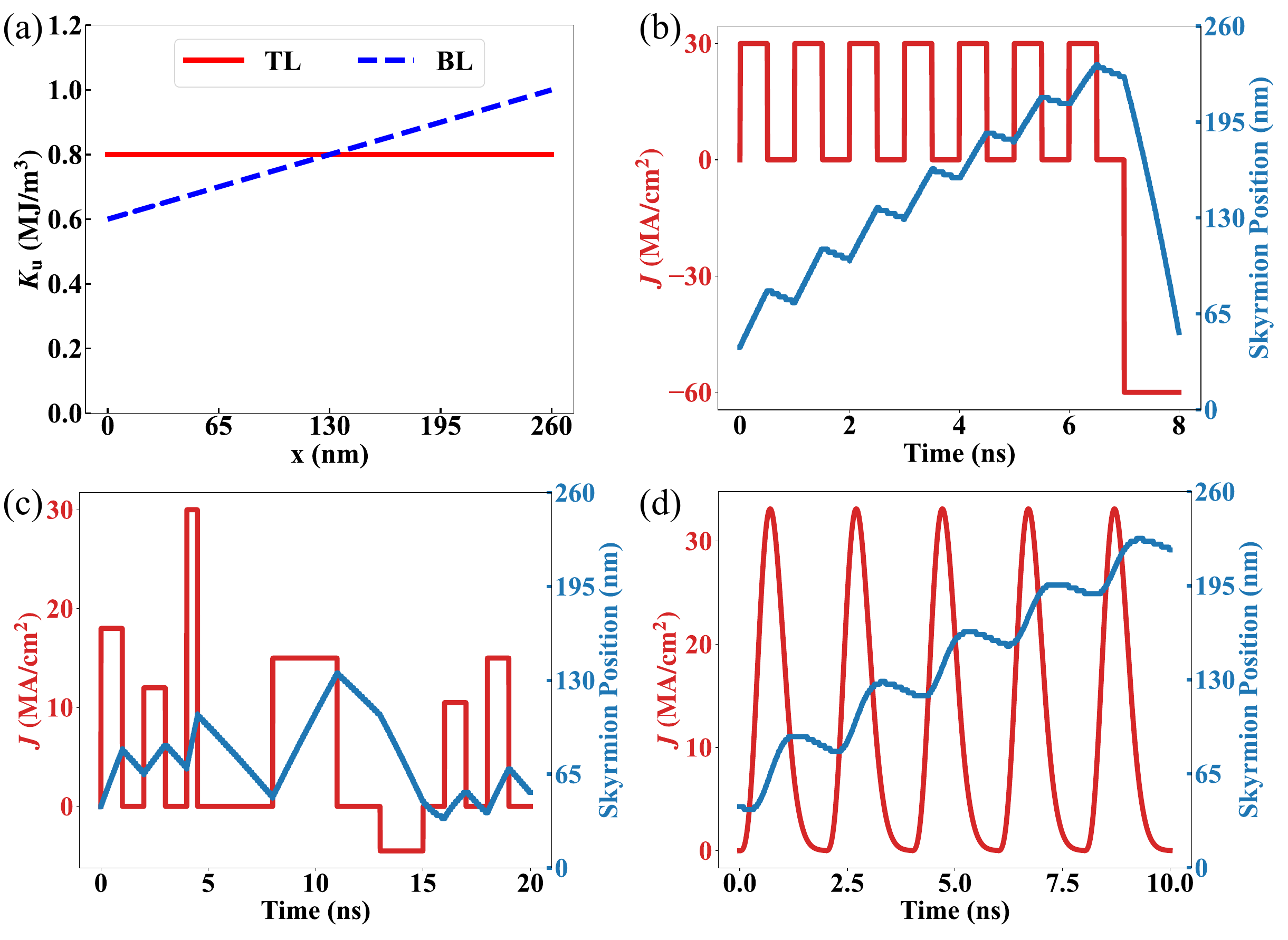}
	\caption{(a) Variation of PMA constant $K_\text{u}$ with position along the length of the nanotrack, for TL and BL. Variation of the applied input current density (red) and corresponding skyrmion position (blue) along nanotrack length with time, for (b) regular square pulse, (c) irregular square pulse, and (d) bio-plausible pulses.}
	\label{Fig:Neuron_Ku}
\end{figure}

It has been reported that the anisotropy gradient can be incorporated by varying the thickness of the FM layer \cite{li2017magnetic}.
However, this method can introduce some complexity into the micromagnetic simulations.
On the other hand, ion bombardment \cite{matczak2014tailoring} during the fabrication of the device is one possible method to incorporate this anisotropy gradient without modifying the thickness of the FM layer.
The gradient can be described by $K_\text{u}(x)=K_\text{u}^{c}+x\Delta{}K_\text{u}$ where $x$ is the position along the length of the nanotrack.
$K_\text{u}^{c}$=0.6$~\mathrm{MJ/m^3}$ and the gradient $\Delta{}K_\text{u}$=1.538 $\mathrm{GJ/m^4}$.
We assume the value of constant $K_\text{u}$ in TL is 0.8$~\mathrm{MJ/m^3}$.
The variation of $K_\text{u}$ along the length of the nanotrack for the FM layer in the monolayer device is shown in Fig.~\ref{Fig:monolayer-bilayer_comparison}(b) whereas the $K_\text{u}$ profile for TL and BL of the bilayer device is shown in Fig.~\ref{Fig:Neuron_Ku}(a).

To simulate the skyrmion motion in monolayer and bilayer devices, we first nucleate a skyrmion (with its core magnetization oriented along $-z$-direction) at $x$ = 40 nm in the FM of the monolayer and also in the TL of the bilayer device.
A skyrmion with +$z$-directed magnetization is also nucleated in the BL of the bilayer device due to the AFM coupling between TL and BL.
To drive the skyrmion, a current pulse of time-period 1~ns and 0.5~ns pulse width is applied in the HM along the +$x$-direction for both types of devices.
The current pulses are applied over 7~ns as shown in Fig.~\ref{Fig:monolayer-bilayer_comparison}(c).
We simulate the monolayer device with current density values, $J$=10, 20, and 30~$\mathrm{MA/cm^2}$, and the bilayer device is simulated only for $J$=30~$\mathrm{MA/cm^2}$.
Skyrmion motion in the monolayer device is also simulated in OOMMF with the parameters described earlier, except $\sigma$.
For the monolayer device, $\sigma$=0 because it consists of only a single FM layer.

The effect of the Magnus force may be observed in the plot of skyrmion position along the lateral dimension of the nanotrack shown in Fig.~\ref{Fig:monolayer-bilayer_comparison}(d).
In this figure, M (B) refers to the monolayer (bilayer) device, and the current density, $J$, is in units of $\mathrm{MA/cm^2}$.
$x$ and $y$ denote the position along length and width of the nanotrack over $x\mbox{-}y$ plane.
When a current is applied along the $+x$-direction in the HM of the monolayer device, two forces act on the skyrmion.
These are the forces due to spin-torque acting along +$x$-direction, accompanied by Magnus force along +$y$-direction.
On the other hand, when no current is applied, the skyrmion feels a force along $-x$-direction due to anisotropy gradient, along with the $-y$-directed Magnus force.
The Magnus force causes the skyrmion to move along a curved path towards the upper edge of the nanotrack due to the direction of the force along +$y$-direction (lower edge when direction of the force is in $-y$-direction), thus showing a non-zero displacement along $y$-direction.
When the current pulse shown in Fig.~\ref{Fig:monolayer-bilayer_comparison}(c) has a magnitude of $J$=10~$\mathrm{MA/cm^2}$, the skyrmion is unable to achieve enough velocity.
Moreover, due to the presence of the Magnus force, the skyrmion moves in a close vicinity around its initial position for the entire pulse train.
Consequently, the skyrmion is unable to reach the right side of the nanotrack.
When the magnitude of the current pulse is increased to $J$=20~$\mathrm{MA/cm^2}$, the skyrmion achieves comparatively higher velocity and it traverses a little further from its initial position along a curved path towards the upper edge of the nanotrack during the high state of the first pulse.
During the low state of the first pulse, the Magnus force acts along $-y$-direction and the skyrmion moves in that direction.
With the rest of the pulses, the skyrmion shows a hopping motion and moves towards the right side of the nanotrack as shown in Fig.~\ref{Fig:monolayer-bilayer_comparison}(d).
Note that as the skyrmion moves deeper into the higher anisotropy region towards the right side, it moves closer to the edge of the nanotrack.
This creates a possibility that the skyrmion would miss the detection unit, and would be unable to fire the neuron.
The problem may worsen for a nanotrack with a longer length.
When the value of $J$ is further increased to 30~$\mathrm{MA/cm^2}$, the skyrmion attains such a high velocity that the Magnus force becomes dominant, moving the skyrmion almost completely in the +$y$-direction to the upper edge of the nanotrack and annihilating it.
(Movie M1-M3 in the Supplemental Materials show the skyrmion motion in the monolayer neuron device for different $J$ values.)

On the other hand, the simulation results for the bilayer device with a square pulse of current density $J$=30~$\mathrm{MA/cm^2}$ shows that the skyrmion moves in the +$x$-direction with negligible displacement in the $y$-direction as shown in Fig.~\ref{Fig:monolayer-bilayer_comparison}(d).
Hence, the skyrmion motion is unaffected by the Magnus force in the bilayer device and the skyrmion will be able to reach the detection unit regardless of the length of the nanotrack and the neuron will fire.
The linear motion of skyrmion in the bilayer device demonstrates its ability to emulate neuron functionality in an effective and well-controlled manner as compared to the monolayer device.
Next, we demonstrate the functionality of our proposed bilayer neuron device for varying input current pulses.

As the skyrmion motion in the bilayer device shows negligible $y$-directed displacement, we investigate the performance of the neuron device by plotting the skyrmion position along the length of the nanotrack versus time along with the corresponding pulse applied to the device.
As described in Sec.~\ref{Sec:device}, the skyrmion motion in the bilayer device with anisotropy gradient within the BL mimics the time integration of the input current pulse and the leak behavior of the biological neuron.
First, is a train of identical 0.5~ns square pulses with $J$=30~$\mathrm{MA/cm^2}$ applied for 7~ns as shown (red color) in Fig.~\ref{Fig:Neuron_Ku}(b).
The skyrmion position along the length of the nanotrack is plotted versus time (blue color) in the same figure.
Once the skyrmion reaches near the end of the nanotrack, the detection unit will detect its presence, fire an output spike and reset the neuron.
Ideally, the neuron should reset itself by moving the skyrmion to the left side of the nanotrack without any external stimulus.
Fig.~\ref{Fig:Leak_Integration}(b) shows that this process takes a longer time ($\approx$23 ns).
To reduce this time, a 1~ns current pulse with $J$=60~$\mathrm{MA/cm^2}$ is applied to the HM in the $-x$-direction to return the skyrmion to its initial position on the left side of the nanotrack as shown in Fig.~ \ref{Fig:Neuron_Ku}(b).

Next, we investigate the response of our device to other types of current pulse trains.
For the irregular square current pulses in Fig.~\ref{Fig:Neuron_Ku}(c), the device is simulated for 20~ns.
The current pulse amplitude, the time between two consecutive pulses, and the current pulse widths are non-uniform.
The simulation result in Fig.~\ref{Fig:Neuron_Ku}(c) shows the initial and final positions of the skyrmion are very near to each other because the irregular input stimulus is insufficient to excite the neuron, and the neuron does not fire within that time window.
\begin{figure*}[t]
	\includegraphics[scale=0.4]{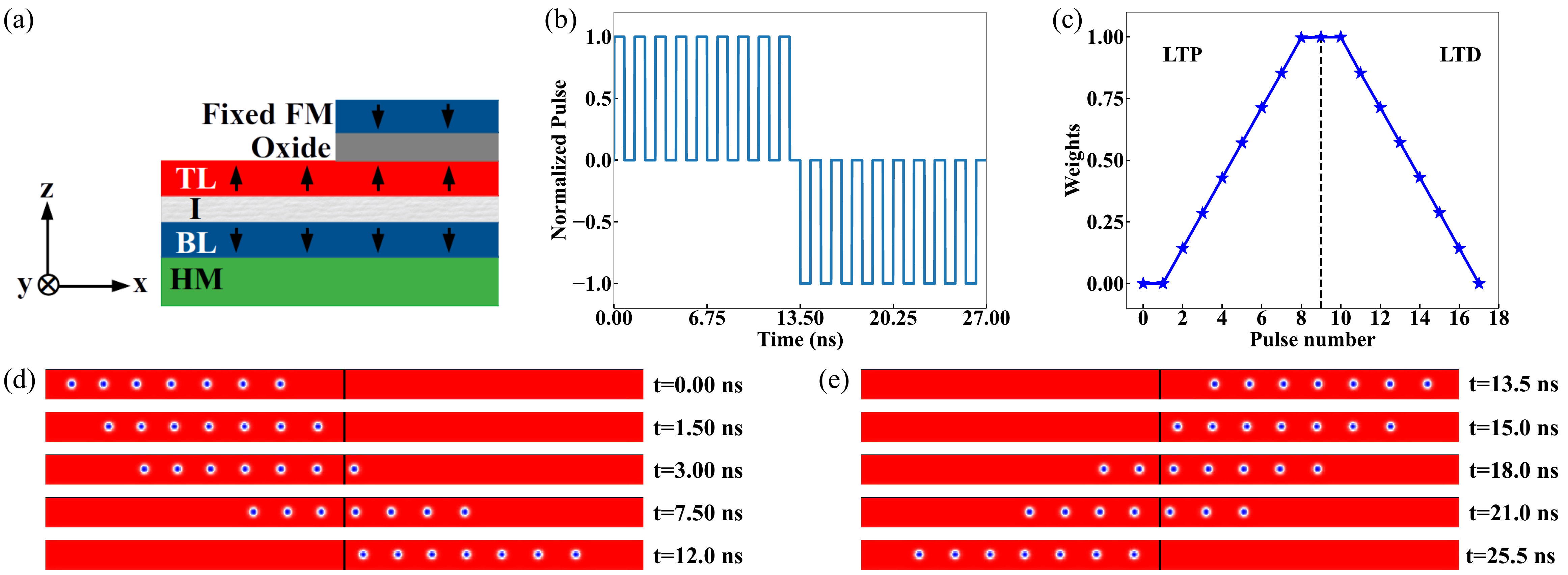}
	\caption{(a) Side-view of the schematic bilayer synapse device. (b) Plot for normalized square pulse train with time, which is given as an input to a synaptic device. (c) The plot of synaptic weight vs. pulse numbers during LTP and LTD. Snapshots of micromagnetic simulation show $m_z$ for TL at different time instants demonstrating (d) LTP and (e) LTD functionality.}
	\label{Fig:Synapse_device}
\end{figure*}

In Fig.~\ref{Fig:Neuron_Ku}(d), we investigate the device response to bio-plausible current pulses (red curve).
Similar to previous results, the device response shows a leaky-integrate functionality and due to the regular occurrence of the pulses, the skyrmion is able to move from the left to right side of the nanotrack as shown (blue color) in Fig.~\ref{Fig:Neuron_Ku}(d).
(Movie M4-M6 in the Supplemental Materials show the skyrmion motion in the bilayer neuron device for different types of pulses.)
The results show that our proposed neuron device can perform the LIF functionality under the influence of various types of pulses, including the bio-plausible pulses produced by realistic biological neurons, but without the accidental skyrmion annihilation that is possible in the monolayer device.

\subsection{The structure as an artificial synapse}\label{Sec:Synapse}

As Fig.~\ref{Fig:Synapse_device}(a) shows, the same bilayer system with a longer nanotrack (1000~nm) and identical width is used to implement the artificial synapse.
As mentioned in Sec.~\ref{Sec:device}, both TL and BL have uniform anisotropy profiles.
Furthermore, the TL nanotrack is divided into two equal portions along its length.
Initially, before any synaptic operations, seven skyrmions are nucleated at the left half of the nanotrack as shown by the first snapshot (t=0~ns) in Fig.~\ref{Fig:Synapse_device}(d), which plots $m_z$ for the TL nanotrack.
These skyrmions can be nucleated by injecting spin-polarized current locally at the desired locations as described in Sec. \ref{Sec:device}.
The nucleation points are chosen to be equidistant to achieve a linear weight update.
Unlike the neuron device, the relaxation time is chosen to be 1~ns to minimize the repulsion force between the skyrmions.
In this figure, a vertical line (black color) marks the border between the left and the right halves for illustration only.
A detector unit, which consists of a fixed FM along with a thin oxide layer forming a magnetic tunnel junction (MTJ) structure as shown in Fig.~\ref{Fig:Synapse_device}(a), is placed over the entire right half of the synapse.
The right half of TL nanotrack works as the free layer and the magnetization of the fixed FM layer in the detector is pinned to the $-z$-direction.
The conductance of the synapse is obtained by calculating the conductance of the MTJ-like structure using~\cite{huang2017magnetic}
\begin{equation}
	G=G_0 \sum_{i}\frac{1+P^2 cos\theta_i}{1+P^2}
\end{equation}  
where $G_0$ is the conductance when all the spins of the free and fixed layer of the detector units are perfectly parallel to each other.
$P$ is the spin polarization, and $\theta_i$ is the angle between the magnetization of the $i$-th cell of the TL right half and the corresponding cell at the fixed FM layer.
In the initial state, when there is no skyrmion in the right half of TL (first snapshot in Fig. \ref{Fig:Synapse_device}(d)), the device has the lowest conductance due to the opposite magnetization of the TL and fixed FM.

When a current is applied along the +$x$-direction in the HM, the skyrmions on the left will move one-by-one into the detector region.
Since the core magnetization of the TL is oriented along the $-z$-direction, the presence of skyrmion in the detector region increases the conductance of the detector unit.
The linear relation between the conductance and the number of skyrmions in the detector unit is shown in Fig.~\ref{Fig:Synapse_Demo}(b).
The conductance value can be converted to weights in a normalized scale using the following relation,
\begin{equation}
	\mathrm{Weight}=\frac{G-G_{min}}{G_{max}-G_{min}}
	\label{Eq:weight}
\end{equation}      
where, $G_{max}$ ($G_{min}$) is the maximum (minimum) conductance for the synapse.

The operation of our proposed synaptic device is investigated using current pulses applied to the HM layer in the $x$-direction.
In this simulation, the magnitude, period and the width for the pulses are $J$=30~$\mathrm{MA/cm^2}$, 1.5~ns and 0.75~ns, respectively.
To emulate the LTP, the current pulses in HM are applied in the $+x$-direction for the first 13.5~ns.
All skyrmions on the left side are driven into the detector region one-by-one.
For LTD, the direction of the current pulse is reversed for the next 13.5~ns, which drives the skyrmions in the opposite direction and out of the detector region one-by-one.
The applied pulse train in a normalized scale is shown in Fig.~\ref{Fig:Synapse_device}(b) and the plot for the increment (decrement) of the synaptic weight with these number of pulses during the LTP (LTD) operation is shown in Fig.~\ref{Fig:Synapse_device}(c).
Snapshots of the $m_{z}$ magnetization component for the TL at different time instants are shown in Fig.~\ref{Fig:Synapse_device}(d) and (e) corresponding to LTP and LTD, respectively.

\begin{figure*}[t]
	\includegraphics[scale=0.28]{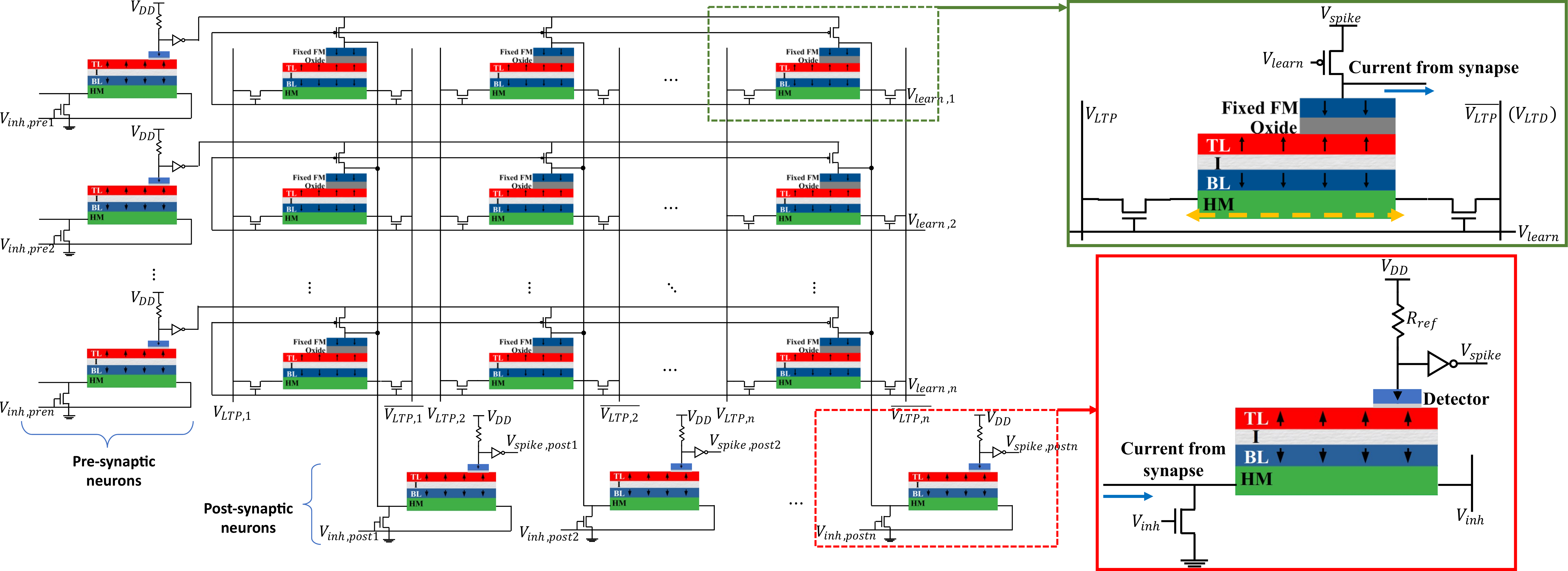}
	\caption{Schematic of the crossbar architecture of skyrmion-based bilayer neuron and synapse devices. A zoomed-in view of the neuron (red dashed box) and synapse (green dashed box) is shown along with the peripheral circuits.}
	\label{Fig:Architecture}
\end{figure*}
The plots in Fig.~\ref{Fig:Synapse_device} demonstrate that the skyrmion motion in the proposed device can achieve different synaptic weight values and that highly linear LTP and LTD operations can be achieved.
The horizontal axis in Fig.~\ref{Fig:Synapse_device}(c) represents the pulse number for the applied current pulse.
Since the period for each pulse is 1.5~ns, pulse number, $n$, corresponds to the time instant $1.5n$~ns in the real time scale.
Consider the LTP operation first.
The first two points in Fig.~\ref{Fig:Synapse_device}(c) (\emph{i.e.}, $n$=0 and 1) show the weight values remain at zero.
This is because, as shown in Fig.~\ref{Fig:Synapse_device}(d), no skyrmion has entered the detector region within the first 1.5~ns of simulation and thus, the device has conductance $G=G_\text{min}$.
At t=3~ns, one skyrmion has entered into the detector region and thus, the weight value increased by one unit.
The weight value for the $n$=2 in Fig.~\ref{Fig:Synapse_device}(c) indicates this increase.
More skyrmions can be shifted into the detector region by applying more current pulses.
Fig.~\ref{Fig:Synapse_device}(c) shows a linear increase of weight value with the pulse number, which is highly desirable for the LTP operation.
At $t$=12~ns, all the seven skyrmions have entered into the detector region and the maximum conductance for the device, $G_\text{max}$ (representing weight value of 1), is achieved. 
Our simulation results show that the first skyrmion moves into the detector region after the second pulse.
If we would have chosen the nucleation points at the positions denoted by the snapshot at t=1.5 ns of Fig. ~\ref{Fig:Synapse_device}(d), then the first skyrmion could have been moved into the detector region just after the first pulse.
This would have removed the delay by shifting the weight update plot in Fig.~\ref{Fig:Synapse_device}(c) to the left but the overall nature of the plot would remain the same.
The first nucleation point was chosen away from the detector as a precaution so that after relaxation, the first skyrmion does not end up inside the detector due to any possible repulsion force.

Let's now consider the LTD operation.
The direction of the current pulse is reversed as shown in Fig.~\ref{Fig:Synapse_device}(b) where the LTD operation begins at $n$=9 (\emph{i.e.}, from $t$=13.5~ns).
Similar to LTP, the synaptic weight remains at 1 for the first two pulses (corresponding to $n$=9 and 10 representing $t$=13.5 and 15~ns, respectively).
This can be explained using Fig.~\ref{Fig:Synapse_device}(e).
Although the current direction is reversed, which moves the skyrmions in the $-x$-direction, no skyrmions leave the detector region until after $t$=15~ns.
Consequently, the device maintains the maximum weight value for the first two pulses.
As more skyrmions exit the detector region with repeated current pulses, the device conductance (weight value) decrease linearly with the number of skyrmions in the detector region.
The linear decrease of the weight value with the pulse number for the LTD operation is shown Fig.~\ref{Fig:Synapse_device}(c), which indicates that 8 distinct states (and hence, 3-bit synaptic device) can be achieved using seven skyrmions.
(Movie M7 in the Supplemental Materials show the skyrmion motion in the bilayer synapse device)

Comparison of LTP and LTD operations in Fig.~\ref{Fig:Synapse_device}(c) also shows that the weight updates for LTP and LTD operations is highly symmetric.
Most other synaptic devices show non-linear and asymmetric LTP and LTD operations, which need to be mitigated by novel design techniques~\cite{fu2019mitigating,chen2015mitigating}.
Such techniques are not needed in our proposed device due to the intrinsicly linear and symmetric LTP and LTD weight update (two highly desirable conditions \cite{xia2019memristive,woo2018resistive,moon2019rram}), which simplifies the circuit design requirements for using our proposed device as an artificial synapse.

To enable a framework to compare the energy efficiencies of different hardware implementations of the artificial neurons/synapses with our proposed devices, the energy consumption of our proposed devices is modeled as follows.
The energy consumption to move the skyrmions in the nanotrack is calculated as,
\begin{equation}
	E=\rho_\text{HM}L_\text{HM}A_\text{HM}J^{2}t_{\mathrm{delay}}
	\label{Eq:E_update}
\end{equation}where $\rho_\text{HM}=100~\mathrm{\mu\Omega-cm}$ is the resistivity of the HM~\cite{nguyen2016spin}.
$L_\text{HM}$, $A_\text{HM}$, $J$, and $t_{\mathrm{delay}}$ are the length (along the $x$-axis), cross-sectional area (in the $yz$-plane) of the HM layer, current density, and time delay to completely move all skyrmions from one half to the other half of the nanotrack (\emph{i.e.}, fully change the synaptic weight between the minimum and maximum values).
Although LTP and LTD operations were simulated for 13.5~ns, the simulations show that 12~ns is required to fully change the synaptic weight.
Thus, $t_{\mathrm{delay}}$ is estimated as 12~ns and $E$=54~fJ.
Since there are a total of seven skyrmions in the TL of our proposed device, the average energy required to adjust the synaptic weight is 7.71~fJ/unit.

\subsection{Neuromorphic architecture}

The synaptic and neuronal behavior of the proposed bilayer neuron and synapse devices have been discussed in the previous sections.
In this section, we present the circuit-level design for a \ac{SNN} hardware accelerator based on an array of the devices.
Fig. \ref{Fig:Architecture} shows the schematic view of the crossbar architecture for offline learning using our proposed skyrmion-based bilayer neuron and synapse devices.
In \ac{SNN}, the neuron firing earliest will inhibit other neurons around it that have also reach their firing thresholds.
In our architecture, inhibition also serves the function of resetting the positions of skyrmions inside the neuron devices since the each neuron needs to be reset after it has fired.
When a neuron is not in the inhibition period, $V_{inh}$ is low. The transistor connected to HM is inactive and the current from synapse will pass through the HM and push the skyrmion from left side to the right side of the device.
When the skyrmion reaches the detector, the resistance at the detector will drop due to the same magnetization direction between the skyrmion core and the detector.
Thus, by choosing an appropriate reference resistance $R_{ref}$, the node voltage between $R_{ref}$ and the detector can be low, which emits a voltage spike at the output of the inverter (\emph{e.g.}, the neuron fires).
After the neuron fires, $V_{inh}$ will be high, which activates the transistor.
This causes a current to flow from the right side of the device to the left side of the device, which pushes the skyrmion back to the initial starting position.
On the other hand, every synapse receives the spike from a previous neuron and supplies current to the next neuron.
The synapse has 2 modes: \emph{spiking} and \emph{learning} modes.
When the synapse is in the spiking mode, $V_{learn}$ is at low voltage. Therefore, the current path between $V_{LTP}$ and $\overline{V_{LTP}}$ is disabled and the current from $V_{spike}$ is passed to the post-synapse neuron.
When the synapse is in the learning mode, $V_{learn}$ is at high voltage and the current path through the HM is enabled (orange dash line). Note that the spike current path is also disabled (blue arrow).
Based on the learning method, $V_{LTP}$ is either at high voltage or low voltage and therefore, the skyrmion train can move either forward or backward.
Therefore, in learning mode, the conductances of the synapse devices can be programmed.
Since long term potentiation (LTP) and long term depression (LTD) are complementary, $\overline{V_{LTP}}$ is the same as $V_{LTD}$.
In the crossbar, $V_{learn, n}$ serves as a row selecting signal and $V_{LTP, n}$ and $\overline{V_{LTP, n}}$ serve as the programming signals.
The input current spikes are applied to the pre-synaptic neurons and the output spikes from pre-synaptic neurons are applied to synapses array.
The current output from these synapses are then applied to post-synaptic neurons.

\begin{figure}[t]
	\includegraphics[scale=0.38]{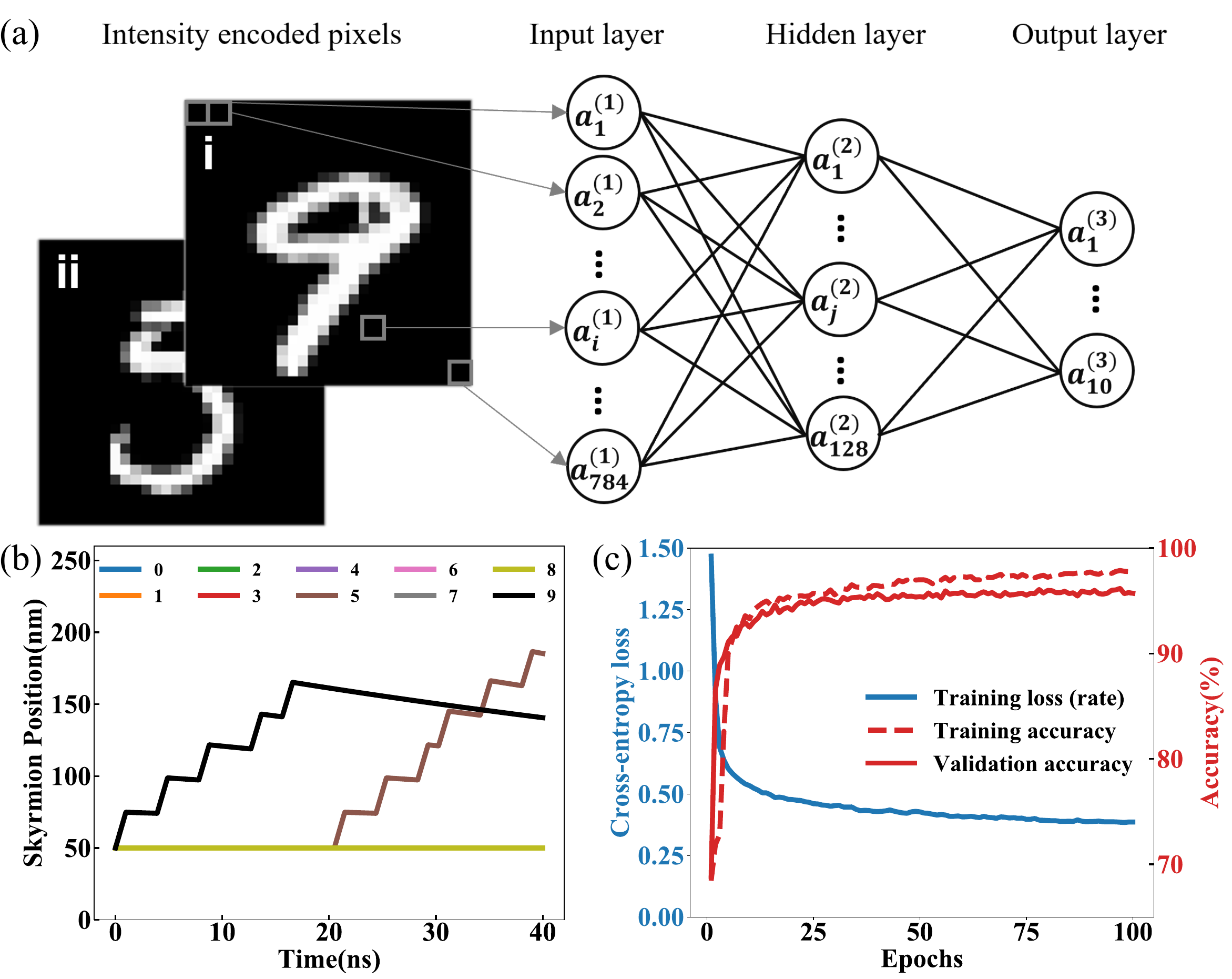}
	\caption{(a) SNN for MNIST handwritten digit recognition. The inset (i) and (ii) are two samples in the MNIST dataset. At the first 20 time steps, (i) is given to the SNN, and after 20 time steps, the input is replaced with (ii). (b) The response of the output layer's neuronal devices to the sample as shown in (i) and (ii). The membrane potential of the LIF neuron is encoded using the skyrmion position of the devices. The neurons at the output layer with the highest membrane potential will be considered as the predicted class of SNN. (c) The training curve of the SNN for 100 epochs, where the training loss is estimated using the rate of the spikes.}
	\label{Fig:SNN}
\end{figure}

\subsection{Demonstration of spiking neural network}

In this section, we evaluate the implementation of a \ac{SNN} neuromorphic circuit based on our proposed bilayer device.
The SNN circuit exploits the linear and symmetric LTP and LTD operations for the offline learning synaptic weight updates, and the intrinsic leaky-integrate behavior for emulating the LIF neuron.
The neurons of the SNN are organized into the $784\times{}128\times{}10$ topology shown in Fig.~\ref{Fig:SNN}(a), and the proposed LIF neurons and 3-bit synapses are trained to perform pattern recognition tasks.
Simulations of the SNN are performed using NengoDL~\cite{rasmussen2019nengodl} and Tensorflow~\cite{tensorflow2015-whitepaper} running on a NVIDIA GeForce RTX 2060 SUPER GPU.

We evaluated the performance of the SNN on the handwritten digit recognition task using the MNIST dataset, which is split into 50000 training samples, 10000 validation samples, and 10000 test samples.
Each sample is reshaped into a column vector corresponding to the input dimension of the SNN.
In our simulation, an amplitude encoding scheme is utilized, where the value of each element in the column vector is encoded based on the intensity of corresponding pixel in the sample image.
By replicating the column vectors for $n$ time steps, the input spike trains are obtained.
For each sample, we run a simulation for 20~ns, where every 2~ns is considered as a time step.
Thus, each sample is converted into a $784\times{}10$ matrix (Fig.~\ref{Fig:SNN}(a,i)).
At each time step, each neuronal device receives a current pulse from the preceding neuron, which move the skyrmions in accordance to:
\begin{equation}
	x_i(t+1) = x_i(t)+\nu\left(\sum_j{w_{ij}i_j(t)}\right)-\eta\left(t\right)
	\label{Eq:V_men}
\end{equation}
where $x_{i}(t)$ is the position of the skyrmion in the $i$-th neuron, $i_{j}(t)$ is the pre-synaptic spike from the $j$-th neuron.
The pre-synaptic spikes are weighted by the synaptic devices ($w_{ij}$).
The skyrmion in the neuron device moves by $\nu$ per unit current and leaks by $\eta$ per time step.
Since the device show remarkable linearity, it is reasonable to utilize the first order Taylor expansion to simulate the integration $\nu(t)=\bar\nu\Delta{t}\sum_j{w_{ij}i_j(t)}$ and leak $\eta=\bar\eta\Delta{t}$, where $\bar\nu$ and $\bar\eta$ is obtained by the first order coefficient of the fitting curve and $\Delta{t}$ is the aforementioned 2~ns time step. 

Two learning methods are commonly used: the conversion-based approach and the spike-based approach~\cite{roy2019towards}.
In this work, conversion-based approach is adopted due to its competitive prediction accuracy~\cite{sengupta2019going, lee2020enabling}.
The main concept of conversion-based learning is to convert the non-differentiable spikes to the differentiable spike rate, which enables the application of a gradient-based backpropagation learning rule to train to SNN.
The correspondence between the synaptic device and backpropagation is as follows.
First, the 8 distinct states of the synaptic device represent the 8 distinct values of a 3-bit synapse.
Second, the continuous position of the skyrmions represents the differentiable full-precision weights.
Once the SNN has been trained in the software, the weight values can be transferred to the synaptic devices in the crossbar array (Fig.~\ref{Fig:Architecture}).
Thereafter, the hardware implementation of the SNN can classify previously unseen data using quantized weights and discrete spikes.

Initially, the skyrmion positions in all the neuron devices are initialized at $x_{i}$=50~nm and the detection units are at $x_{i}$=150~nm.
At each time step, neuron devices that receive input current pulses will move their skyrmions according to Eq.~(\ref{Eq:V_men}).
Note that the current density of the input spikes is $J=25J_\text{sp}~\mathrm{MA/cm^2}$, where $J_\text{sp}$ is the amplitude of the spikes.
Once the skyrmions reach the detection units, the neurons will emit pre-synaptic spikes and reset the position of their skyrmions.
The neuron devices that do not receive current pulses will leak according Eq.~(\ref{Eq:V_men}) due to the $K_u$ gradient.
The neurons at the final layers fire less frequently due to sparser spike activity in the deeper layers.
Hence, the neuron that receives the earliest spike in the output layer (\emph{i.e.}, the neuron with the skyrmion farthest from the initialized position) will be considered as the predicted class for to the input presented to the SNN.
Fig.~\ref{Fig:SNN}(b) shows the skyrmion positions in the output neurons when the input shown in Fig.~\ref{Fig:SNN}(a,i) is presented to the SNN, which the SNN  recognized as the digit `9'.
To test the response of the skyrmions, after 20~ns, we change the input to Fig.~\ref{Fig:SNN}(a,ii). 
We notice that the skyrmion position of the neuron corresponds to digit `9' leaks due to the $K_{u}$ gradient, and the neuron corresponds to digit `5' will respond to the new stimuli.
Typically, we also care about energy consumption of the neuromorphic system based on the abovementioned mechanism.
We evaluate all the images in test set and utilized Eq.~\eqref{Eq:E_update} to estimate that each neuron consumes an energy of 1.07~fJ per time step on average.

The \ac{SNN} is trained for 100 epochs using the Adam optimizer~\cite{kingma2014adam} with a mini-batch size of 500. Fig.~\ref{Fig:SNN}(c) shows that the training loss is always decreasing.
However, the \ac{SNN} model tends to overfit to the training sample during the latter epochs.
The highest validation accuracy is achieved at epoch~91, where the classification accuracy on the test set is 96.23\% with an estimated latency of around 40~ns.
Notably, latency is also an important metric for \ac{SNN}, where in this case we can improve by reducing the time steps.
If the number of time steps is too low, the output layer may not obtain sufficient information, which results in a poor prediction.
Hence, to obtain a high prediction accuracy as well as an acceptable latency, the number of time steps needs to be optimized.
As we reduce the time steps from 20 to 4, 93.55~\% of the test images can still be predicted correctly within a latency of $\sim$8~ns.
This represents $\sim{}$80\% improvement in latency with increase in prediction errors from 3.77\% to 6.45\%.

\section{Conclusion}

In this paper, we proposed a skyrmion-based bilayer device concept that can be used as artificial neurons and synapses. The advantages of the proposed device concept are that it (i) does not suffer from the unwanted Magnus force, (ii) achieves highly linear and symmetric LTP and LTD synaptic weight update operations, and (iii) can be engineered to emulate a synapse or a LIF neuron. The leak functionality of the neuron is achieved by incorporating an anisotropy gradient into the device. Using micromagnetic simulations, we showed that the Magnus force degrades the functionality of the monolayer neuron device whereas our proposed bilayer neuron device is unaffected. We also showed that removing the anisotropy gradient allows the bilayer device to be used as a highly linear synapse having symmetric weight updates during LTP and LTD operations. Evaluation of an SNN implemented using our proposed neuron and synapse devices was performed. The SNN was trained using a modified spike-based backpropagation technique to perform an image recognition task and achieved an accuracy of 96.23\% when classifying the MNIST handwritten digits dataset.

\bibliography{References}

\end{document}